\documentclass[twoside]{article}

\usepackage{fancyhdr}
\usepackage{graphicx}

\newcommand{\myfigpatha}{.}
\newcommand{\myfigpathb}{.}
\newcommand{\myfigsize}{0.45}
\newcommand{\myfigsizeb}{0.3}
\newcommand{\myfigsizec}{0.8}
\newcommand{\myfigsized}{0.65}

\newcommand{\myint}[2]{\int\limits_{#1}^{#2}}

\newcommand{\mybpar}[1]{\left( #1 \right)}
\newcommand{\beq}{\begin{equation}}
\newcommand{\eeq}{\end{equation}}
\newcommand{\beqa}{\begin{eqnarray}}
\newcommand{\eeqa}{\end{eqnarray}}
\newcommand{\mymat}[2]{\left( \begin{array}{*{#1}{c}} #2 \end{array} \right) }

\begin{document}
\fancypagestyle{plain}{
	\fancyhf{}
        \fancyhf[HCO]{To appear in 
``Nanoscience, Engineering and Technology Handbook'' edited by William Goddard,
Donald Brenner, Sergey Lyshevski and Gerald Iafrate, published by CRC Press.}
}
\thispagestyle{plain}
\pagestyle{plain}

\title{Resistance of a Molecule} 
\author{Magnus Paulsson\thanks{mpaulsso@purdue.edu}, Ferdows Zahid\thanks{zahidf@purdue.edu} and 
Supriyo Datta\thanks{datta@purdue.edu} \\
School Of Electrical \& Computer Engineering, \\
Purdue University, \\
1285 Electrical Engineering Building,\\ 
West Lafayette, IN - 47907-1285, USA} 
\date{\today} 
\maketitle
\thispagestyle{plain}
\pagestyle{plain}

\vspace{-1cm}
\input{TOC.toc}

\clearpage

\fancypagestyle{myfancy}{
\fancyhead{}
\fancyhead[RO,LE]{\thepage}
\fancyhead[RE]{Resistance of a Molecule}
\fancyhead[LO]{\leftmark}
\renewcommand{\headrulewidth}{1pt}
\fancyfoot{}
}
\pagestyle{myfancy}

\section{Introduction}
\label{sect.intro}

In recent years, several experimental groups have reported measurements of the current-voltage 
(I-V) characteristics of individual or small numbers of molecules. Even 
three-terminal measurements showing evidence of transistor action has been
reported using carbon nanotubes \cite{avouris.s01,bachtold.s01} as well as self-assembled 
monolayers of conjugated polymers \cite{schon.n01}. 
These developments have attracted much attention from the semiconductor 
industry and there is great interest from
an applied point of view to model and understand the capabilities of molecular conductors.
At the same time, this is also a topic of great interest from the point of view of basic physics.
A molecule represents a quantum dot, at least an order of magnitude smaller than semiconductor 
quantum dots, which allows us to study many of the same mesoscopic and/or many-body effects
at far higher temperatures.

\begin{figure}[!htb]
\begin{center}
\includegraphics[width=\myfigsizec \columnwidth]{\myfigpatha/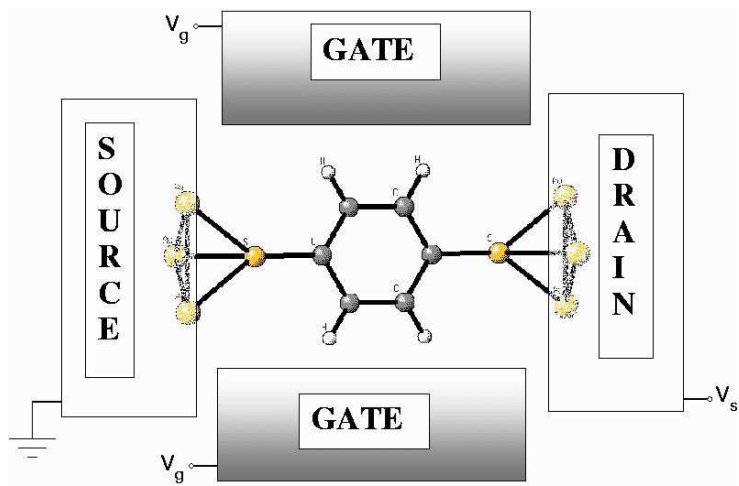}
\end{center}
\caption{Conceptual picture of a ``molecular transistor'' showing a short molecule 
(Phenyl dithiol, PDT) sandwiched between source and drain contacts. Most experiments so far
lack good contacts and do not incorporate the gate electrodes.
} 
\label{fig.first}
\end{figure}

\begin{figure}[!htb]
\begin{center}
\includegraphics[width=\myfigsizec \columnwidth]{\myfigpatha/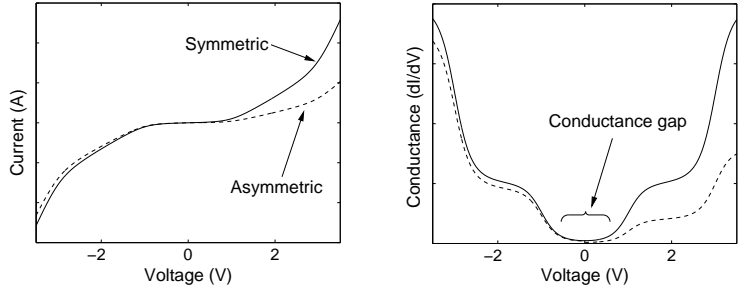}
\end{center}
\caption{Schematic picture, showing general properties of measured current-voltage 
(I-V) and conductance (G-V) characteristics for molecular wires.
Solid line, symmetrical I-V. Dashed line, asymmetrical I-V. }
\label{fig.i-v}
\end{figure}

So what is the resistance of a molecule? More specifically,
what do we see when we connect a short molecule between two metallic contacts as shown in Fig.~\ref{fig.first} and
measure the current (I) as a function of the voltage (V)? Most commonly we get 
I-V characteristics of the type sketched in Fig.~\ref{fig.i-v}. This has been observed
using many different approaches including breakjunctions 
\cite{andres.s96,reed.s97,joachim.prb99,kergueris.nano99,reichert.unpub01}, 
scanning probes \cite{datta.sup00,rosink.prb00,joachim.cpl97,datta.jcp98},
nanopores \cite{chen.apl00} and a host of other methods (see for example \cite{porath.n00}).
A number of theoretical models have been developed for calculating the I-V characteristics
of molecular wires using semi-empirical \cite{datta.jcp98,magoga.prb99,hall.jcp00,
paulsson.prb01,emberly.prb00} as well as first principles 
\cite{damle.prb01,taylor.prb01,lang.prl00,damle.cp02,xue.jcp01,palacios.prb01,seminario.jpca99} theory.

\begin{figure}[!htb]
\begin{center}
\includegraphics[width= \columnwidth]{\myfigpatha/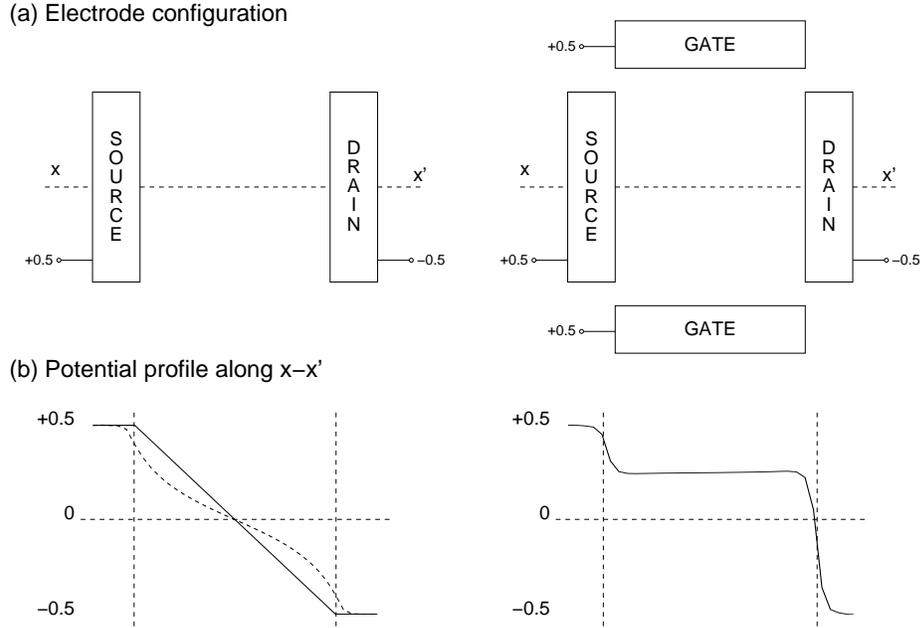}
\end{center}
\caption{Schematic picture, showing potential profile for two geometries, 
without gate (left) and with gate (right).}
\label{fig.profile}
\end{figure}

Our purpose in this chapter is to provide an intuitive explanation for the observed I-V characteristics
using simple models to illustrate the basic physics. However, it should be noted that molecular electronics is a 
fast developing field and much of the excitement arises from the possibility of discovering novel
physics beyond the paradigms discussed here. To cite a simple example, very few experiments to date \cite{schon.n01}
incorporate the gate electrode shown in Fig.~\ref{fig.first}, and we will largely ignore the gate in this chapter. 
However,
the gate electrode can play a significant role in shaping the I-V characteristics and deserves more attention.
This is easily appreciated by looking at the applied potential profile $U_{app}$ generated by the electrodes in the 
absence of the molecule. This potential profile satisfies the Laplace equation without any net charge anywhere, 
and is obtained by solving:
\beq
\nabla \cdot \mybpar{\epsilon \nabla U_{app}}=0 
\eeq
subject to the appropriate boundary values on the electrodes (Fig.~\ref{fig.profile}). It is
apparent that the electrode geometry has a significant influence on the potential profile that it 
imposes on the molecular species and this in term could obviously affect the I-V characteristics in a
significant way.
After all, it is well known that a three-terminal metal/oxide/semiconductor Field Effect Transistor 
(MOSFET) with a gate electrode has 
a very different I-V characteristic compared to a two terminal ``n-i-n'' diode: The current in a MOSFET 
saturates under increasing bias, but the current in an ``n-i-n'' diode keeps increasing indefinitely.
In contrast to the MOSFET, whose I-V is largely dominated by classical electrostatics, the I-V 
characteristics of molecules is determined by a more interesting interplay between
nineteenth century physics (electrostatics) and twentieth century physics (quantum transport) and it is
important to do justice to both aspects.

\emph{Outline of chapter}: We will start in section \ref{sect.quality} with a qualitative
discussion of the main factors affecting the I-V characteristics of molecular conductors, using
a simple toy model to illustrate their role. However, this 
toy model misses two important factors: (1) Shift in the energy level 
due to {\emph{charging effects}} as the molecule loses or gains electrons and (2)
{\emph{broadening}} of the energy levels due to their finite lifetime
arising from the coupling ($\Gamma_1$ and $\Gamma_2$) to the two
contacts. Once we incorporate these effects (section \ref{sect.charging}) 
we obtain more realistic I-V plots, even though the toy model assumes that conduction takes place independently
through individual molecular levels. In general, however, multiple energy levels are simultaneously involved
in the conduction process.
In section \ref{sect.theory} we will describe the non-equilibrium
Green's function (NEGF) formalism which can be viewed as a generalized
version of the one-level model to include multiple levels or conduction
channels. This formalism provides a convenient framework for describing quantum transport \cite{datta.sup00b} and can 
be used in conjunction with ab initio or semi-empirical Hamiltonians as described in a set of related articles 
\cite{avik.chap02,zahid.mark02}. Here (section \ref{sect.results}) we will illustrate the NEGF formalism with a simple 
semi-empirical model for a gold wire, '$n$' atoms long and one atom in cross-section. We could call this a $\mbox{Au}_n$ 
molecule through that is not how one normally thinks of a gold wire. However, this example is particularly instructive 
because it shows the lowest possible ``resistance of a molecule'' per channel, which is 
$\pi \hbar/e^2=12.9 \;\mbox{k}\Omega$ \cite{datta.meso95}.

\section{Qualitative Discussion}
\label{sect.quality}

\subsection{Where is the Fermi energy?}
\begin{figure}[!htb]
\begin{center}
\includegraphics[width=\myfigsizec \columnwidth]{\myfigpathb/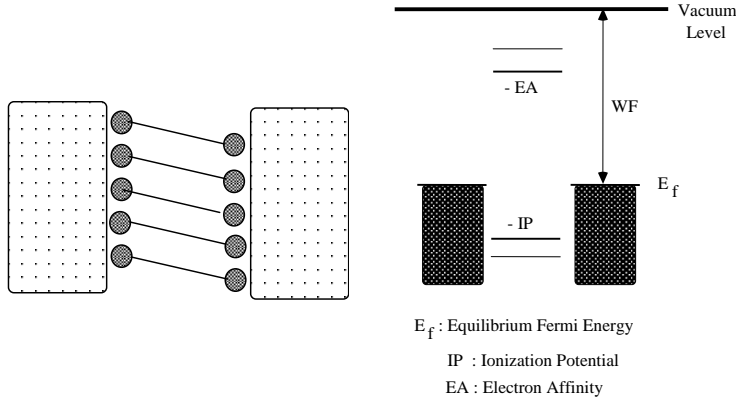}
\end{center}
\caption{Equilibrium energy level diagram for a metal-molecule-metal sandwich for 
a weakly coupled molecule.}
\label{fig.datta1a}
\end{figure}

\emph{Energy Level Diagram}:  
The first step in understanding the current (I) vs.~voltage (V) curve for a molecular 
conductor is to draw an energy level diagram and locate the Fermi energy. Consider 
first a molecule sandwiched between two metallic contacts, but with very weak electronic 
coupling. We could then line up the energy levels as shown in Fig.~\ref{fig.datta1a} 
using the metallic  
work function (WF) and the electronic affinity ($EA$) and ionization potential ($IP$) of the 
molecule. For example, a (111) gold surface has a work function of $\sim 5.3$ eV while the electron 
affinity and ionization potential, $EA_0$ and $IP_0$, for isolated phenyl dithiol (Fig.~\ref{fig.first}) in the gas 
phase have been reported to be $\sim 2.4$ eV and $8.3$ eV respectively \cite{lias.jopcrd88}. 
These values are associated with electron emission and injection to and from a vacuum and may 
need some modification to account for the metallic contacts. For example the actual $EA$, $IP$
will possibly be modified from $EA_0$, $IP_0$ due to the image potential $W_{im}$ associated 
with the metallic contacts \cite{yellow.book}:
\beqa
EA & = & EA_0 + W_{im}	\\	
IP & = & IP_0 - W_{im}
\eeqa

The probability of the molecule losing an electron to form a positive ion is equal to 
$e^{(WF-IP)/k_B T}$ while the probability of the molecule gaining an electron to form a negative 
ion is equal to $e^{(EA-WF)/k_B T}$. We thus expect the molecule to remain neutral as 
long as both $(IP - WF)$ and $(WF - EA)$ are much larger than $k_B T$, a condition that 
is usually satisfied for most metal-molecule combinations. 
Since it costs too much energy to transfer one electron into or out of the molecule, 
it prefers to remain neutral in equilibrium. 

\begin{figure}[!htb]
\begin{center}
\includegraphics[width=\myfigsizec \columnwidth]{\myfigpathb/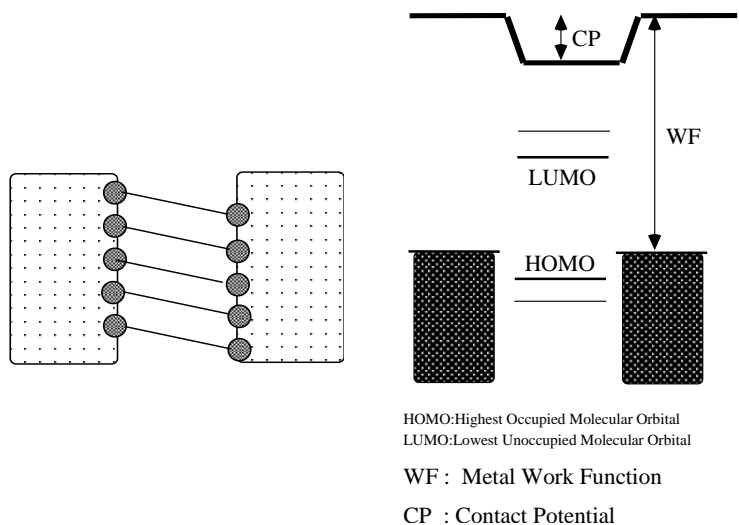}
\end{center}
\caption{Equilibrium energy level diagram for a metal-molecule-metal sandwich for 
a molecule strongly coupled to the contacts.}
\label{fig.datta1b}
\end{figure}

The picture changes qualitatively if the 
molecule is chemisorbed directly on the metallic contact (Fig.~\ref{fig.datta1b}). 
The molecular energy 
levels are now broadened significantly by the strong hybridization with the delocalized 
metallic wavefunctions, making it possible to transfer fractional amounts of charge 
to or from the molecule.
Indeed there is a charge transfer which causes a change in the electrostatic potential
inside the molecule and the energy levels of the molecule
are shifted by a contact potential (CP), as shown.

It is now more appropriate to describe transport in terms 
of the HOMO-LUMO levels 
associated with incremental charge transfer \cite{parr.dft89} 
rather than the affinity and ionization levels associated with integer charge transfer. 
Whether the molecule-metal coupling is strong enough for this to occur depends on the 
relative magnitudes of the single electron charging energy ($U$) and energy level 
broadening ($\Gamma$). As a rule of thumb, 
if $U >> \Gamma$, we can expect the structure to be in the Coulomb Blockade (CB) 
regime characterized by integer charge transfer; otherwise it is in the 
self-consistent field (SCF) regime characterized by fractional charge transfer. 
This is basically the same criterion that one uses for the Mott transition in periodic structures, 
with $\Gamma$ playing the role of the hopping matrix element. It is important to note 
that for a structure to be in the CB regime both contacts must be weakly coupled, since 
the total broadening $\Gamma$ is the sum of the individual broadening due to the two contacts. 
Even if only one of the contacts is coupled strongly we can expect $\Gamma \sim U$ 
thus putting the structure in the SCF regime. Fig.~\ref{fig.datta6} in Sec.~\ref{sect.toymodel2} 
illustrates the I-V characteristics in the CB regime using a toy model. However, a moderate 
amount of broadening destroys this effect (see Fig.~\ref{fig.toynocb}) and in this chapter we will
generally assume that the conduction is in the SCF regime.

\emph{Location of the Fermi energy:} The location of the Fermi energy relative to
the HOMO and LUMO levels is probably the most important factor in determining
the current (I) versus voltage (V) characteristics of molecular conductors.
Usually it lies somewhere inside the HOMO-LUMO gap. 
To see this, we first note that $E_f$ is located by the requirement 
that the number of states below the Fermi energy must be equal to 
the number of electrons in the molecule. 
But this number need not be equal to the integer number we expect for a neutral molecule. 
A molecule does not remain exactly neutral when connected to the contacts. 
It can and does pick up a fractional charge depending on the work function of the metal. 
However, the charge transferred ($\delta n$) for most metal-molecule combinations is usually 
much less than one. If $\delta n$ were equal to +1, the Fermi energy would lie on the 
LUMO while if $\delta n$ were -1, it would lie on the HOMO. Clearly for values in between, 
it should lie somewhere in the HOMO-LUMO gap.

A number of authors have performed detailed calculations to locate the Fermi energy 
with respect to the molecular levels for a phenyl dithiol molecule sandwiched between 
gold contacts, but there is considerable disagreement.
Different theoretical groups have placed it close to the LUMO \cite{hall.jcp00,lang.prl00} 
or to the HOMO \cite{datta.jcp98,damle.prb01}.
The density of states inside the HOMO-LUMO 
gap is quite small making the precise location of the Fermi energy very sensitive to 
small amounts of electron transfer, a fact that could have a significant effect on both
theory and experiment. 
As such it seems justifiable to treat $E_f$ as a ``fitting parameter'' within 
reasonable limits when trying to explain experimental I-V curves.

\begin{figure}[!htb]
\begin{center}
\includegraphics[width=\myfigsized \columnwidth]{\myfigpatha/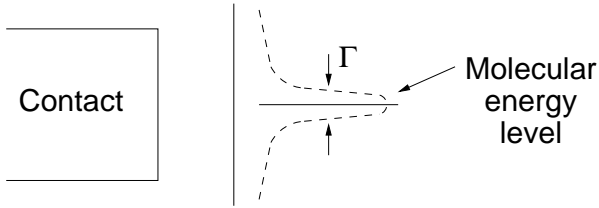}
\end{center}
\caption{Energy level broadening.}
\label{fig.broad}
\end{figure}

\emph{Broadening by the contacts}:
``Common sense'' suggests that the strength of coupling of the molecule to the contacts is 
important in determining the current flow - the stronger the coupling, the larger the current.
A useful quantitative measure of the coupling is the resulting broadening $\Gamma$ of the molecular
energy levels, see Fig.~\ref{fig.broad}. 
This broadening $\Gamma$ can also be related to the time $\tau$
it takes for an electron placed in that level to escape into the contact: $\Gamma=\hbar / \tau$.
In general, the broadening $\Gamma$ could be 
different for different energy levels. Also it is convenient to define two quantities $\Gamma_1$
and $\Gamma_2$, one for each contact, with the total broadening $\Gamma=\Gamma_1+\Gamma_2$.

One subtle point. Suppose an energy level is located well below the Fermi energy in the contact,
so that the electrons are prevented from escaping by the exclusion principle. Would $\Gamma$ be zero?
No, the broadening would still be $\Gamma$, independent of the degree of filling of the contact
as discussed in Ref.~\cite{datta.sup00b}. This observation is implicit in the NEGF formalism, though
we do not invoke it explicitly.

\subsection{Current flow as a ``balancing act''} 

\begin{figure}[!htb]
\begin{center}
\includegraphics[width=\myfigsizec \columnwidth]{\myfigpathb/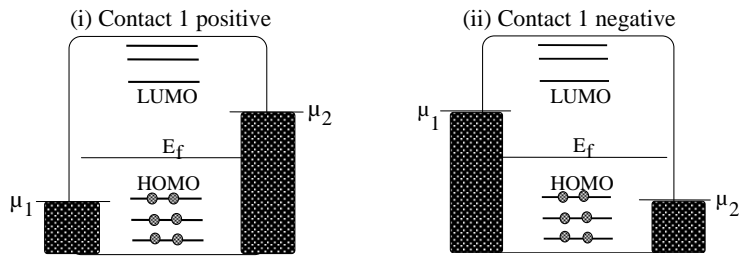}
\end{center}
\caption{Schematic energy level diagram of metal-molecule-metal structure when contact 1 is
(i) positively biased and when contact 1 is (ii) negatively biased with respect to contact 2.}
\label{fig.datta2}
\end{figure}

Once we have drawn an equilibrium energy level diagram, we can understand the 
process of current flow which involves a non-equilibrium situation where the 
different reservoirs (e.g., the source and the drain) have different electrochemical 
potentials $\mu$, see Fig.~\ref{fig.datta2}. For example if a positive voltage V is applied externally to the 
drain with respect to the source, then the drain has an electrochemical potential 
lower than that of the source by $e V$: $\mu_2=\mu_1-e V$. The source and drain 
contacts thus have different Fermi functions and each seeks to bring the active 
device into equilibrium with itself. The source keeps pumping electrons into it 
hoping to establish equilibrium. But equilibrium is never achieved as the drain 
keeps pulling electrons out in its bid to establish equilibrium with itself. 
The device is thus forced into a balancing act between two reservoirs with 
different agendas which sends it into a non-equilibrium state intermediate between 
what the source and drain would like to see. 
To describe this balancing process we need a kinetic equation that keeps track 
of the in and out-flow of electrons from each of the reservoirs.

\begin{figure}[!htb]
\begin{center}
\includegraphics[width=\myfigsized\columnwidth]{\myfigpathb/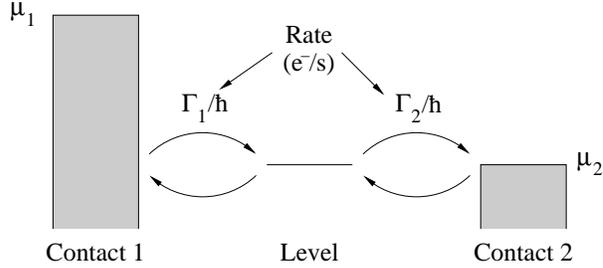}
\end{center}
\caption{Illustration of the kinetic equation.}
\label{fig.seq}
\end{figure}

\emph{Kinetic equation}: This balancing act is easy to see if we consider a simple one level system,
biased such that the energy $\epsilon$ lies in between the electrochemical potentials of the two
contacts (Fig.~\ref{fig.seq}). An electron in this level can escape into contacts 1 and 2 at a rate 
of $\Gamma_1/\hbar$ and $\Gamma_2/\hbar$ respectively. If the level were in equilibrium with contact 
1 then the number of electrons occupying the level would be given by:
\beq
N_1=2 (\mbox{for spin}) f(\epsilon,\mu_1)
\eeq
where 
\beq
f(\epsilon,\mu)=\frac{1}{1+e^{\frac{\epsilon-\mu}{k_B T}}}
\eeq
is the Fermi function.
Similarly if the level were in equilibrium with contact 2 the number would be:
\beq
N_2=2 (\mbox{for spin}) f(\epsilon,\mu_2) 
\eeq

Under non-equilibrium conditions the number of electrons N will be somewhere in between 
$N_1$ and $N_2$. To determine this number we write a steady state \emph{kinetic equation} that
equates the net current at the left junction:
\beq
I_L=\frac{e \Gamma_1}{\hbar}(N_1-N) \label{eq.currL}
\eeq 
to the net current at the right junction:
\beq
I_R=\frac{e \Gamma_2}{\hbar}(N-N_2) \label{eq.currR}
\eeq 
Steady state requires $I_L=I_R$, from which we obtain:
\beq
N=2 \frac{\Gamma_1 f(\epsilon,\mu_1)+\Gamma_2 f(\epsilon,\mu_2)}{\Gamma_1+\Gamma_2} \label{eq.avikN}
\eeq
so that from Eq.~\ref{eq.currL} or \ref{eq.currR} we obtain the current:
\beq
I=\frac{2 e}{\hbar} \frac{\Gamma_1 \Gamma_2 }{\Gamma_1+\Gamma_2}
\mybpar{f(\epsilon,\mu_1)-f(\epsilon,\mu_2)} \label{eq.currTot}
\eeq

Eq.~\ref{eq.currTot} follows very simply from an elementary model, but it serves to
illustrate a basic fact about the process of current flow. No current will flow if 
$f(\epsilon,\mu_1)=f(\epsilon,\mu_2)$. A level that is way below both electrochemical potentials
$\mu_1$ and $\mu_2$ will have $f(\epsilon,\mu_1)=f(\epsilon,\mu_2)=1$ and will not contribute to the 
current, just like a level that is way above both potentials $\mu_1$ and $\mu_2$ and has 
$f(\epsilon,\mu_1)=f(\epsilon,\mu_2)=0$. It is only when the level lies in between $\mu_1$ and $\mu_2$
(or within a few $k_B T$ of $\mu_1$ and $\mu_2$) 
that we have $f(\epsilon,\mu_1) \ne f(\epsilon,\mu_2)$ and a current flows.
Current flow is thus the result of the \emph{difference in opinion} between the contacts.
One contact would like to see more electrons (than $N$) occupy the level and keeps pumping them in,
while the other would like to see fewer than $N$ electrons and keeps pulling them out.
The net effect is a continuous transfer of electrons from one contact to another.

\begin{figure}[!htb]
\begin{center}
\includegraphics[width=\myfigsize \columnwidth]{\myfigpatha/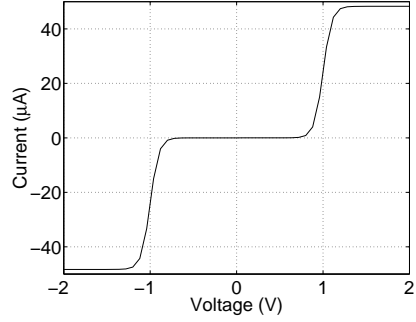}
\end{center}
\caption{The current-voltage (I-V) characteristics for our toy model with $\mu_1=E_f-e V/2$, $\mu_2=E_f+e V/2$,
$E_f=-5.0$ eV, $\epsilon_0=-5.5$ eV and $\Gamma_1=\Gamma_2=0.2$ eV. 
Matlab code in appendix {\protect \ref{app.toy1}} ($U=0$).} 
\label{fig.toy0}
\end{figure}

Fig.~\ref{fig.toy0} shows a typical current ($I$) versus voltage ($V$) calculated from Eq.~\ref{eq.currTot},
using the parameters indicated in the caption. At first the current is zero because both $\mu_1$, 
$\mu_2$ are above the energy level. 
\begin{center}
\includegraphics[width=\myfigsizeb \columnwidth]{\myfigpatha/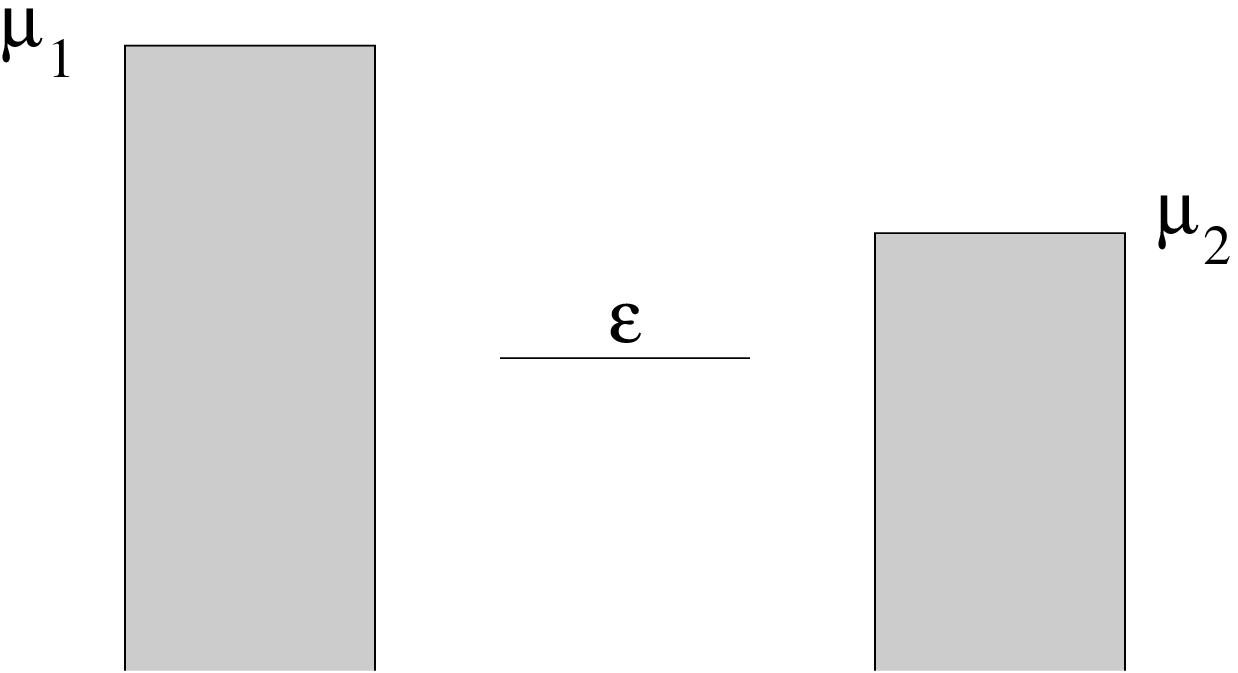}
\end{center}
Once $\mu_2$ drops below the energy level, the current increases 
to $I_{max}$ which is the maximum current that can flow through one level and is obtained from 
Eq.~\ref{eq.currTot} by setting $f(\epsilon,\mu_1)=1$ and $f(\epsilon,\mu_2)=0$:
\beq
I_{max}=\frac{2 e}{\hbar} \Gamma_{\mbox{eff}}=\frac{2 e}{\hbar} \frac{\Gamma_1 \Gamma_2}{\Gamma_1+\Gamma_2}
\eeq 
\begin{center}
\includegraphics[width=\myfigsizeb \columnwidth]{\myfigpatha/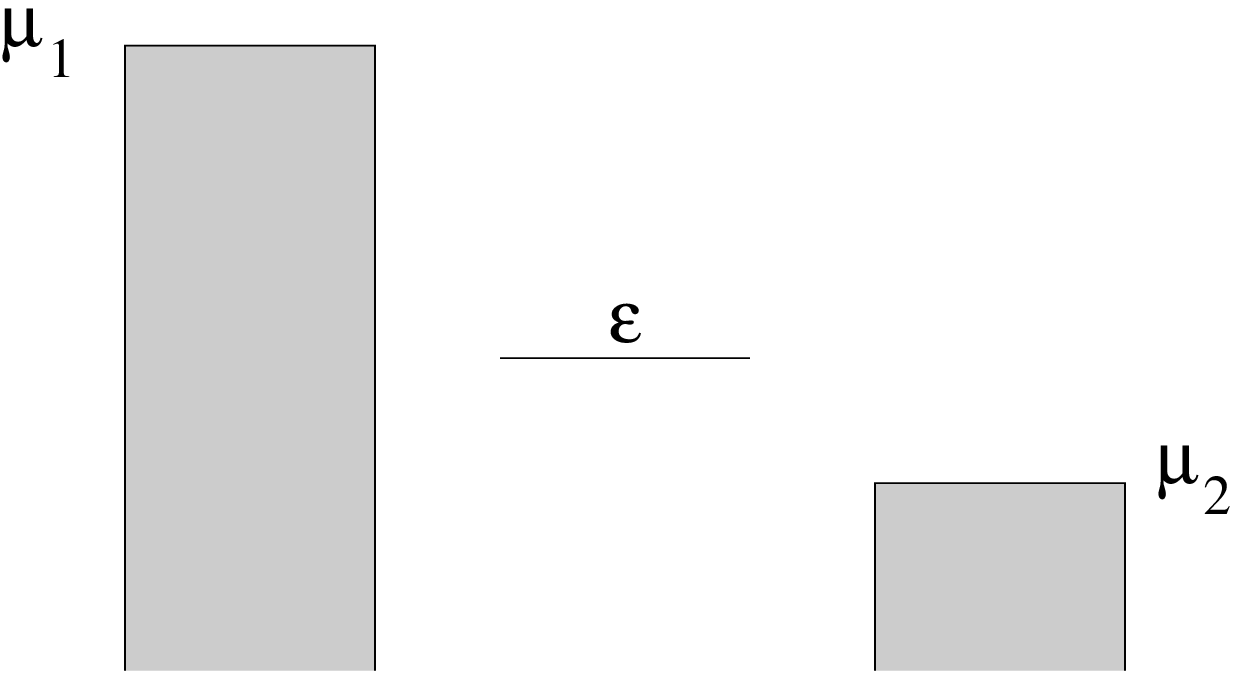}
\end{center}

Note that in Fig.~\ref{fig.toy0} we have set $\mu_1=E_f-e V/2$ and $\mu_2=E_f+e V/2$. We 
could, of coarse, just as well have set $\mu_1=E_f-e V$ and $\mu_2=E_f$. But the, the 
average potential in the molecule would be $-V/2$ and we would need to shift $\epsilon$
appropriately. It is more convenient to choose our reference such that the average molecular 
potential is zero and there is no need to shift $\epsilon$.
 
Note that the current is proportional to $\Gamma_{\mbox{eff}}$ which is the parallel combination of $\Gamma_1$ 
and $\Gamma_2$. This seems quite reasonable if we recognize that $\Gamma_1$ and $\Gamma_2$ represents 
the strength of the coupling to the two contacts and as such are like two \emph{conductances in series}.
For long conductors we would expect a third conductance in series representing the actual conductor.
This is what we usually have in mind when we speak of conductance. But short conductors
have virtually zero resistance and what we measure is
essentially the contact or interface resistance\footnote{
Four-terminal measurments have been used to separate the contact from the device resistance 
(see for example Ref. \cite{rStormer}).}. This is an important
conceptual issue, that caused much argument and controversy in the
1980's and was finally resolved when experimentalists measured the
conductance of very short conductors and found it approximately equal
to $2e^2/h$ which is a fundamental constant equal to 77.8 $\mu$A/V. The
inverse of this conductance $h/2e^2 = 12.9$ k$\Omega$  is now believed
to represent the minimum contact resistance that can be achieved for a
one-channel conductor.  Even a copper wire with a one atom cross section
will have a resistance at least this large. 
Our simple one-level model (Fig,~\ref{fig.toy0}) does not predict this result because
we have treated the level as discrete, but the more complete treatment
in later sections will show it.

\section{Coulomb Blockade?}
\label{sect.charging}
As we mentioned in section \ref{sect.quality}, a basic question we need to answer is 
whether the process of conduction through the molecule belongs to the Coulomb Blockade (CB) or
the Self-Consistent Field (SCF) regime. In this section, we will first discuss a simple model 
for charging effects (section \ref{sect.restricted}) and then look at the distinction between 
the simple SCF regime and the CB regime (section \ref{sect.toymodel2}). Finally,
in section \ref{sect.broadening} we show how moderate amount of level broadening often 
destroy CB effects making a simple SCF treatment quite accurate.

\subsection{Charging Effects}
\label{sect.restricted}
\begin{figure}[!htb]
\begin{center}
\includegraphics[width=\myfigsizeb \columnwidth]{\myfigpathb/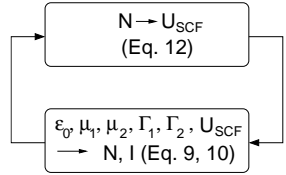}
\end{center}
\caption{Illustration of the SCF procedure.} 
\label{fig.scf}
\end{figure}

Given the level ($\epsilon$), broadening ($\Gamma_1$, $\Gamma_2$) and the 
electrochemical potentials $\mu_1$ and $\mu_2$ of the two contacts, we can
solve Eq.~\ref{eq.currTot} for the current $I$.
But we want to include charging effects in the calculations. 
Therefore, we add a potential $U_{SCF}$ due to the 
change in the number of electrons from the equilibrium value ($f_0=f(\epsilon_0,E_f)$):
\beq
U_{SCF}= U \mybpar{N-2 f_0} \label{eq.uscf}
\eeq 
similar to a Hubbard model.
We then let the level $\epsilon$ float up or down by this potential:
\beq
\epsilon=\epsilon_0+U_{SCF} \label{eq.datta12}
\eeq
Since the potential depends on the number of electrons, we 
need to calculate the potential using the self consistent procedure shown in Fig.~\ref{fig.scf}.

\begin{figure}[!htb]
\begin{center}
\includegraphics[width=\myfigsizec \columnwidth]{\myfigpathb/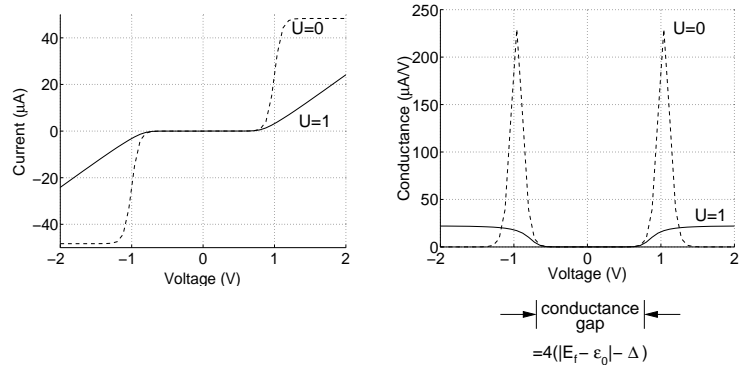}
\end{center}
\caption{The current-voltage (I-V) characteristics (left) and conductance-voltage (G-V) (right) 
for our toy model with $E_f=-5.0$ eV, $\epsilon_0=-5.5$ eV and $\Gamma_1=\Gamma_2=0.2$ eV.
Solid lines, charging effects included ($U=1.0$ eV). Dashed line, no charging ($U=0$). 
Matlab code in appendix {\protect \ref{app.toy1}}
} 
\label{fig.datta4}
\end{figure}

Once the converged solution is obtained, the current is calculated from Eq.~\ref{eq.currTot}.
This very simple model captures much of the observed  physics of molecular conduction. 
For example, the results obtained by 
setting  $E_f=-5.0$ eV, $\epsilon_0=-5.5$ eV, $\Gamma_1= 0.2$ eV, 
$\Gamma_2=0.2$ eV are shown in Fig.~\ref{fig.datta4} with ($U = 1.0$ eV) and 
without ($U = 0$ eV) charging effects. 
The finite width of the conductance peak (with $U=0$) 
is due to the temperature used in the calculations
($k_B T=0.025$ eV). 
Note how the inclusion of charging tends to broaden the sharp peaks in conductance, 
even though we have not included any extra level broadening in this calculation.
The size of the conductance gap is directly
related to the energy difference between the molecular energy level and the Fermi energy.
The current starts to increase when the voltage reaches $1$ V, which is exactly
$2\left|E_f-\epsilon_0\right|$ as would be expected even from a theory with no charging.
Charging enters the picture only at
higher voltages, when a chemical potential tries to cross the level.
The energy level shifts in energy (Eq.~\ref{eq.datta12}) if the charging energy is non-zero.
Thus, for a small charging energy, the chemical potential easily crosses the level giving a
sharp increase of the current. If the charging energy is large, the current increases gradually 
since the energy level follows the chemical potential due to the charging.

\begin{figure}[!htb]
\begin{center}
\includegraphics[width=\myfigsized \columnwidth]{\myfigpathb/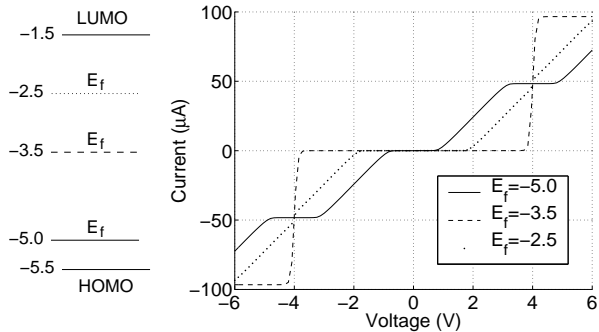}
\end{center}
\caption{Right, the current-voltage (I-V) characteristics for the two level 
toy model for three different values of the Fermi energy ($E_f$).
Left, the two energy levels (LUMO$=-1.5$ eV, HOMO$=-5.5$ eV) and the three different Fermi energies 
($-2.5$, $-3.5$, $-5.0$) used in the calculations. 
(Other parameters used $U=1.0$ eV, $\Gamma_1=\Gamma_2=0.2$ eV)
Matlab code in appendix {\protect \ref{app.toy2}}
} 
\label{fig.twolevel}
\end{figure}

{\it What determines the conductance gap?} The above discussion
shows that the conductance gap is equal to $4 \mybpar{\left| E_f-\epsilon_0 \right|-\Delta}$ 
where $\Delta$ is equal to $\sim 4 k_B T$ (plus $\Gamma_1+\Gamma_2$ if broadening is included, 
see section \ref{sect.broadening}), and 
$\epsilon_0$ is the HOMO or LUMO 
level whichever is closest to the Fermi energy, as pointed out in Ref.~\cite{datta.prl97}.
This is unappreciated by many who associate the conductance 
gap with the HOMO-LUMO gap. However, we believe that what conductance measurements show 
is the gap between the Fermi energy and the nearest molecular level\footnote{
With very asymmetric contacs, the conductance gap could be equal to the HOMO-LUMO gap as
commonly assumed in interpreting STM spectra. However, we belive that the picture presented here
is more accurate unless the contact is so strongly coupled that there is a significant density of
Metal-Induced Gap States (MIGS)\cite{zahid.mark02}}.
Fig.~\ref{fig.twolevel} shows the I-V characteristics calculated using
a two-level model (obtained by a straightforward extension of the one-level model) with
the Fermi energy located differently within the HOMO-LUMO gap giving different conductance gaps corresponding to 
the different values of $\left| E_f-\epsilon_0 \right|$. Note that with the Fermi energy
located halfway in between, the conductance gap is twice the HOMO-LUMO gap and the I-V shows no evidence 
of charging effects because the depletion of the HOMO is neutralized by the charging of the LUMO. This
perfect compensation is unlikely in practice, since the two levels will not couple identically to the contacts
as assumed in the model.   

\begin{figure}[!htb]
\begin{center}
\includegraphics[width=\myfigsizec \columnwidth]{\myfigpathb/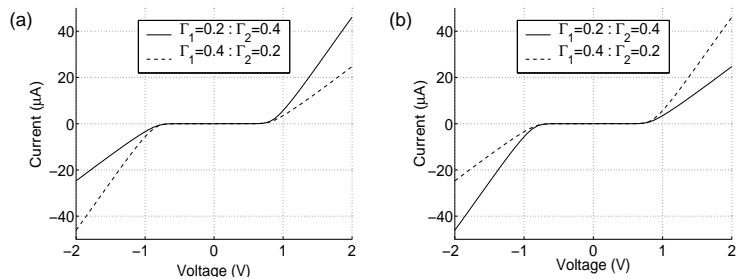}
\end{center}
\caption{The current-voltage (I-V) characteristics 
for our toy model ($E_f=-5.0$ eV and $U=1.0$ eV).
(a) Conduction through HOMO ($E_f>\epsilon_0=-5.5$ eV). 
(b) Conduction through LUMO ($E_f<\epsilon_0=-4.5$ eV). 
Solid lines, $\Gamma_1=0.2$ eV $<$  $\Gamma_2=0.4$ eV.
Dashed lines, $\Gamma_1=0.4$ eV $>$  $\Gamma_2=0.2$ eV.
Matlab code in appendix {\protect \ref{app.toy1}}. Here positive voltage 
is defined as a voltage that lowers the chemical potential of contact 1.} 
\label{fig.datta5}
\end{figure}

A very interesting effect that can be observed is the asymmetry of the 
I-V characteristics if $\Gamma_1 \ne \Gamma_2$ as shown in Fig.~\ref{fig.datta5}. 
This may explain several experimental results which show asymmetric I-V 
\cite{datta.jcp98,joachim.prb99} as discussed by Ghosh {\it et al.} \cite{avik.unpub01}.
Assuming that the current is conducted through the HOMO level ($E_f>\epsilon_0$), 
the current is less when a positive voltage is applied to the strongly coupled contact, 
see Fig.~\ref{fig.datta5}(a). This is due to the effects of charging as has been 
discussed in more detail in Ref.~\cite{avik.unpub01}. 
Ghosh {\it et al.} also shows that this result will reverse if the conduction is through the LUMO level. 
We can simulate this situation by setting $\epsilon_0$ equal to $-4.5$ eV, $0.5$
eV above the equilibrium Fermi energy $E_f$. 
The sense of asymmetry is now reversed as shown in Fig.~\ref{fig.datta5}(b). 
The current is larger when a positive voltage is applied to the strongly coupled contact.
Comparing with STM measurements seems to favor the first case, i.e., conduction through
the HOMO \cite{avik.unpub01}.

\subsection{Unrestricted Model} 
\label{sect.toymodel2}

\begin{figure}[!htb]
\begin{center}
\includegraphics[width=\myfigsize \columnwidth]{\myfigpathb/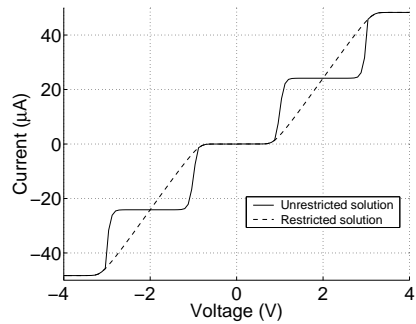}
\end{center}
\caption{The current-voltage (I-V) characteristics for restricted (dashed line) and 
unrestricted solutions (solid line).
$E_f=-5.0$ eV, $\epsilon_0=-5.5$ eV, $\Gamma_1=\Gamma_2=0.2$ eV and $U=1.0$ eV.
Matlab code in appendix {\protect \ref{app.toy1} and \ref{app.toy4}}
} 
\label{fig.datta6}
\end{figure}

In the previous examples (Figs.~\ref{fig.datta4}, \ref{fig.datta5}) 
we have used values of $\Gamma_{1,2}$ that are smaller than the 
charging energy $U$. However, under these conditions 
one can expect Coulomb Blockade (CB) effects which are not captured by a 
``restricted solution'' which assumes that both spin orbitals see the same 
self-consistent field. However, an unrestricted solution, which allows the 
spin degeneracy to be lifted, will show these effects\footnote{
The unrestricted one-particle picture discussed here provides at least a reasonable
qualitative picture of CB effects, though a complete description requires a more advanced many
particle picture \cite{cb85}. The one-particle picture leads to one of many possible states 
of the device depending on our initial guess, while a full many particle 
picture would include all states.}. 
For example, if we replace Eq.~\ref{eq.datta12} with ($f_0=f(\epsilon_0,E_f)$):
\beqa
\epsilon_\uparrow=\epsilon_0+U\mybpar{N_\downarrow-f_0}\label{eq.spinup}\\
\epsilon_\downarrow=\epsilon_0+U\mybpar{N_\uparrow-f_0}\label{eq.spindown}
\eeqa
where the up-spin level feels a potential due to the down-spin electrons 
and vice-versa, then we obtain I-V curves as shown in Fig.~\ref{fig.datta6}.

If the SCF iteration is started with a spin degenerate solution, 
the same restricted solution as before is obtained. 
However, if the iteration is started with a spin non-degenerate solution 
a different looking I-V is obtained. 
The electrons only interact with the the electron of the opposite
spin. Therefore, the chemical potential of one contact can cross one energy level of the
molecule since the charging of that level only affects the opposite spin level. 
Thus, the I-V contains two separate steps separated by $U$ instead of a single 
step broadened by $U$.

For a molecule chemically bonded to a metallic surface, e.g., a PDT molecule bonded by a
thiol group to a gold surface, the broadening $\Gamma$ is expected to be of the same magnitude 
or larger than $U$. This washes out CB effects as shown in Fig.~\ref{fig.toynocb}. 
Therefore, the CB is not expected in this case. However, if the coupling to both
contacts is weak we should keep the possibility of CB and the importance of 
unrestricted solutions in mind.

\subsection{Broadening}
\label{sect.broadening}
\begin{figure}[!htb]
\begin{center}
\includegraphics[width=\myfigsize \columnwidth]{\myfigpathb/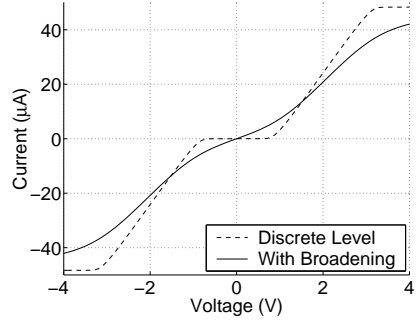}
\end{center}
\caption{The current-voltage (I-V) characteristics:
Solid line, include broadening of the level by the contacts.
Dashed line, no broadening, same as solid line in Fig.~\ref{fig.datta4}.
Matlab code in appendix {\protect \ref{app.toy3} and \ref{app.toy1}}
($E_f=-5.0$, $\epsilon_0=-5.5$, $U=1$ and $\Gamma_1=\Gamma_2=0.2$ eV).}
\label{fig.datta7}
\end{figure}

So far we have treated the level $\epsilon$ as discrete,
ignoring the broadening $\Gamma=\Gamma_1+\Gamma_2$ that accompanies the 
coupling to the contacts. To take this into account we need to replace the discrete
level with a Lorentzian density of states $D(E)$:
\beq
D(E)=\frac{1}{2 \pi} \frac{\Gamma}{(E-\epsilon)^2+(\Gamma/2)^2}
\eeq
As we will see later, $\Gamma$ is in general \emph{energy-dependent} so that $D(E)$ can deviate significantly
from a Lorentzian shape.
We modify Eqs.~\ref{eq.avikN}, \ref{eq.currTot} for $N$ and $I$ to include an integration
over energy:
\beqa
N&=& 2 \myint{-\infty}{\infty}\mbox{d}E D(E) \,
\frac{\Gamma_1 f(E,\mu_1)+\Gamma_2 f(E,\mu_2)}{\Gamma_1 +\Gamma_2 } \label{eq.datta23}\\
I&=& \frac{2 e}{\hbar} \myint{-\infty}{\infty}\mbox{d}E D(E) \,
\frac{\Gamma_1 \Gamma_2}{\Gamma_1+\Gamma_2} \mybpar{f(E,\mu_1)-f(E,\mu_2)}
\eeqa

\begin{figure}[!htb]
\begin{center}
\includegraphics[width=\myfigsize \columnwidth]{\myfigpatha/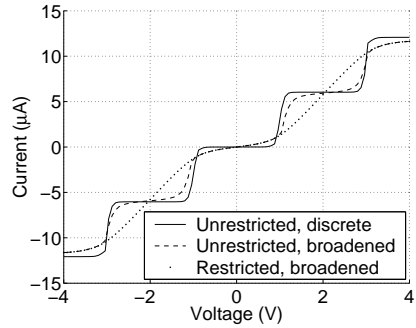}
\end{center}
\caption{Current-voltage (I-V) characteristics showing the Coulomb blockade:
discrete unrestricted model (solid line, Matlab code in appendix \protect\ref{app.toy4}) 
and the broadened unrestricted model (dashed line, \protect\ref{app.toy5}).
The dotted line shows the broadened restricted model without Coulomb blockade ({\protect \ref{app.toy3}}). 
For all curves the following parameters were used  $E_f=-5.0$, $\epsilon_0=-5.5$, $U=1$ and $\Gamma_1=\Gamma_2=0.05$ eV.} 
\label{fig.toycb}
\end{figure}
The charging effect is included as before by letting the center $\epsilon$, of 
the molecular density of states, float up or down according to Eqs.~(\ref{eq.uscf},\ref{eq.datta12}) for the 
restricted model or Eqs.~(\ref{eq.spinup},\ref{eq.spindown}) for the unrestricted model.

\begin{figure}[!htb]
\begin{center}
\includegraphics[width=\myfigsize \columnwidth]{\myfigpatha/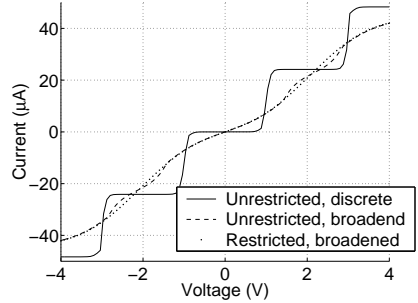}
\end{center}
\caption{Current-voltage (I-V) characteristics showing the suppression of the Coulomb blockade by broadening:  
discrete unrestricted model (solid line) 
and the broadened unrestricted model (dashed line).
The dotted line shows the broadened restricted model. 
$\Gamma_1=\Gamma_2=0.2$ eV.}
\label{fig.toynocb}
\end{figure}

For the  restricted model, the only effect of broadening is to smear out the I-V characteristics as
evident from Fig.~\ref{fig.datta7}. The same is true for the unrestricted model as long as the 
broadening is much smaller than the charging energy (Fig.~\ref{fig.toycb}). But moderate amounts of 
broadening can destroy the Coulomb blockade effects completely and make the I-V characteristics look 
identical to the restricted model (Fig.~\ref{fig.toynocb}).
With this in mind, we will use the restricted model in the remainder of this chapter.

\section{Non Equilibrium Green's Function (NEGF) Formalism}
\label{sect.theory}

The one-level toy model described in the last section includes the three basic
factors that influence molecular conduction, namely, $E_f-\epsilon_0$, $\Gamma_{1,2}$ and $U$.
However, real molecules typically have multiple levels that often broaden and overlap in energy.
Note that the two-level model (Fig.~\ref{fig.twolevel}) in the last section treated
the two levels as \emph{independent} and such models can be used only if the levels do not overlap.
In general we need a formalism that can do justice to multiple levels with arbitrary
broadening and overlap. The non-equilibrium Green's function (NEGF) formalism described in this section 
does just that.

In the last section we obtained equations for the number of electrons, 
$N$ and the current, $I$ for a one-level model with broadening. 
It is useful to rewrite these equations
in terms of the Green's function $G(E)$ which is defined as follows:
\beq
G(E)=\mybpar{E-\epsilon+i\frac{\Gamma_1+\Gamma_2}{2}}^{-1}
\eeq
The density of states $D(E)$ is proportional to the spectral function $A(E)$ defined as:
\beqa
A(E)&=&-2 \mbox{Im} \left\{ G(E) \right\} \\
D(E)&=&\frac{A(E)}{2 \pi}    
\eeqa 
while the number of electrons, $N$ and the current, $I$ can be written as:
\beqa
N&=& \frac{2}{2\pi} \myint{-\infty}{\infty}\mbox{d}E \, \mybpar{ 
{\left| G(E)\right|^2 \Gamma_1 f(E,\mu_1)+\left| G(E)\right|^2 \Gamma_2 
f(E,\mu_2)}} \label{eq.datta55}\\
I&=& \frac{2 e}{h} \myint{-\infty}{\infty}\mbox{d}E \, 
\Gamma_1 \Gamma_2\left| G(E)\right|^2 \mybpar{f(E,\mu_1)-f(E,\mu_2)} \label{eq.datta56}
\eeqa

In the NEGF formalism the single energy level $\epsilon$ is replaced 
by a Hamiltonian matrix $\left[H\right]$ while the broadening $\Gamma_{1,2}$ is replaced 
by a complex energy-dependent self-energy matrix $\left[\Sigma_{1,2}(E)\right]$ so that
the Green's function becomes a matrix given by:
\beq
G(E)=\mybpar{E S-H-\Sigma_1-\Sigma_2}^{-1}
\label{eq.datta60}
\eeq
where $S$ is the identity matrix of the same size as the other matrices and
the broadening matrices $\Gamma_{1,2}$ are defined as the imaginary
(more correctly as the anti-Hermitian) parts of $\Sigma_{1,2}$:
\beq
\Gamma_{1,2}=i \mybpar{\Sigma_{1,2}-\Sigma_{1,2}^\dagger}
\eeq
The spectral function is the anti-Hermitian part of the Green's function:
\beq
A(E)=i \mybpar{G(E)-G^\dagger(E)}
\eeq
from which the density of states $D(E)$ can be calculated by taking the trace:
\beq
D(E)=\frac{\mbox{Tr}\mybpar{A S}}{2 \pi} \label{eq.dosfroma}
\eeq
The density matrix $\left[\rho\right]$
is given by, c.f., Eq.~\ref{eq.datta55}:
\beq
\rho= \frac{1}{2 \pi} \myint{-\infty}{\infty} \left[ f(E,\mu_1) G \Gamma_1 G^\dagger  
+ f(E,\mu_2) G \Gamma_2 G^\dagger \right] \, \mbox{d}E 
\label{eq.rho}
\eeq
from which the total number of electrons, N can be calculated by taking a trace:
\beq
N=\mbox{Tr}\mybpar{\rho S} \label{eq.nfromrho}
\eeq
The current is given by, c.f., Eq.~\ref{eq.datta56}:
\beq
I=\frac{2e}{h} \myint{-\infty}{\infty} \left[ {\mbox{Tr} \mybpar{\Gamma_1 G \Gamma_2 G^\dagger} 
\mybpar{f(E,\mu_1)-f(E,\mu_2)}} \right] \, \mbox{d}E \label{eq.datta61}
\eeq

Equations \ref{eq.datta60} through \ref{eq.datta61} constitute the basic equations of the
NEGF formalism which have to be solved self consistently with a suitable scheme
to calculate the self-consistent potential matrix $\left[U_{SCF}\right]$, c.f., Eq.~\ref{eq.datta12}:
\beq
H=H_{0}+U_{SCF}
\eeq
where $H_0$ is the bare Hamiltonian (like $\epsilon_0$ in the toy model)
and $U_{SCF}$ is an appropriate functional of the density 
matrix $\rho$:
\beq
U_{SCF}=F\left(\rho\right) \label{eq.datta70}
\eeq
This self-consistent procedure is essentially the same as in Fig.~\ref{fig.scf} for
the one level toy model, except that scalar quantities have been replaced by matrices:
\beqa
\epsilon_0 &\rightarrow &\left[H_0\right] \\
\Gamma &\rightarrow &\left[\Gamma\right], \left[\Sigma\right] \\
N &\rightarrow &\left[\rho\right] \\
U_{SCF} &\rightarrow &\left[U_{SCF}\right] 
\eeqa
The sizes of all these matrices is ($n\times n$), $n$ being the 
number of basis functions used to describe the \emph{molecule}. 
Even the self-energy matrices $\Sigma_{1,2}$ are of
this size although they represent the effect of infinitely large
contacts. In the remainder of this section and the next section, 
we will describe the procedure used to evaluate the Hamiltonian matrix $H$,
the self-energy matrices $\Sigma_{1,2}$ and the 
functional ``$F$'' used to evaluate the self-consistent potential
$U_{SCF}$ (see Eq.~\ref{eq.datta70}). But the point to 
note is that once we know how to evaluate these matrices, Eqs.~\ref{eq.datta60} 
through \ref{eq.datta70} can be used straight forwardly to calculate the current.

\emph{Non-orthogonal basis}:
The matrices appearing above depend on the basis functions that we use.
Many of the formulations in quantum chemistry use non-orthogonal
basis functions and the matrix equations \ref{eq.datta60} through \ref{eq.datta70}
are still valid as is, except that the elements of the matrix
$\left[S\right]$ in Eq.~\ref{eq.datta60} represents the overlap of
the basis function $\phi_m(\bar{r})$:
\beq
S_{mn}=\myint{}{}\mbox{d}^3r \, \phi_m^*(\bar{r})\phi_n(\bar{r})
\eeq
For orthogonal bases, $S_{mn}=\delta_{mn}$ so that $S$ is the 
identity matrix as stated earlier. The fact that the matrix equations \ref{eq.datta60}
through \ref{eq.datta70} are valid even in a non-orthogonal representation is
not self-evident and is discussed in Ref.~\cite{zahid.mark02}.

\emph{Incoherent Scattering}: One last comment about
the general formalism. The formalism as described 
above neglects all incoherent scattering processes inside the molecule.
In this form it is essentially equivalent to the Landauer formalism \cite{bagwell.prb89}.
Indeed our expression for the current (Eq.~\ref{eq.datta61}) is exactly
the same as in the transmission formalism with the transmission $T$ given by
$\mbox{Tr}\mybpar{\Gamma_1 G \Gamma_2 G^\dagger}$. But it should be noted that,
the real power of the NEGF formalism lies in its ability to provide a
first principles description of incoherent scattering processes - something
we do not address in this chapter and leave for future work.

\emph{A practical consideration}: Both Eq.~\ref{eq.rho} and \ref{eq.datta61}
require an integral over all energy. This is not a problem in Eq.~\ref{eq.datta61}
because the integrand is non-zero only over a limited range where $f(E,\mu_1)$ differs
significantly from $f(E,\mu_2)$. But in Eq.~\ref{eq.rho} the integrand is non-zero over a large 
energy range and often has sharp structures making it numerically challenging to evaluate the 
integral. One way to address this problem is to write:
\beq
\rho=\rho_{eq}+\Delta \rho
\eeq
where $\rho_{eq}$ is the equilibrium density matrix given by:
\beq
\rho= \frac{1}{2 \pi} \myint{-\infty}{\infty} f(E,\mu) \left[ G \Gamma_1 G^\dagger  
+ G \Gamma_2 G^\dagger \right] \, \mbox{d}E 
\label{eq.rhoeq}
\eeq
and $\Delta \rho$ is the change in the density matrix under bias:
\beq
\rho= \frac{1}{2 \pi} \myint{-\infty}{\infty} G \Gamma_1 G^\dagger \left[ f(E,\mu_1)-f(E,\mu) \right]   
+ G \Gamma_2 G^\dagger \left[ f(E,\mu_2)-f(E,\mu) \right]\, \mbox{d}E 
\label{eq.rhodelta}
\eeq
The integrand in Eq.~\ref{eq.rhodelta} for $\Delta \rho$ is non-zero only
over a limited range (like Eq.~\ref{eq.datta61} for $I$) and is evaluated relatively easily.
The evaluation of $\rho_{eq}$ (Eq.~\ref{eq.rhoeq}) however still
has the same problem but this integral (unlike the original Eq.~\ref{eq.rho})
can be tackled by taking advantage of the method of contour integration as described in 
Ref.~\cite{brandbyge.prb02,zeller.ssc82}.

\section{An Example: Quantum Point Contact (QPC)}
\label{sect.results}

\begin{figure}[!htb]
\begin{center}
\includegraphics[width=\myfigsizec \columnwidth]{\myfigpatha/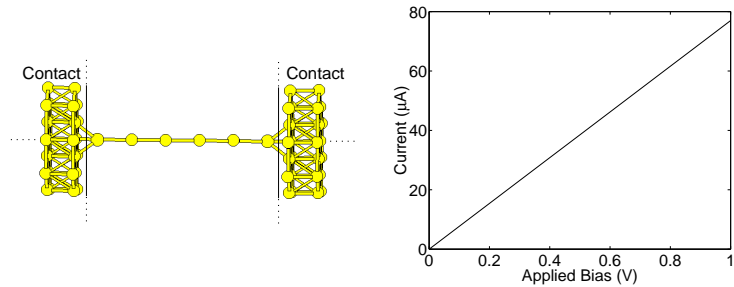}
\end{center}
\caption{Left, wire consisting of six gold atoms forming a Quantum Point Contact
 (QPC). 
Right, quantized conductance ($I=\frac{e^2}{\pi \hbar} V$).}
\label{fig.qpc1}
\end{figure}

Consider for example a gold wire stretched between two gold surfaces as shown in Fig.~\ref{fig.qpc1}.
One of the seminal results of mesoscopic physics is that such a wire has a quantized conductance 
equal to $\frac{e^2}{\pi \hbar}\sim 77.5 \; \mu \mbox{A}/\mbox{V}\sim \mybpar{12.9\; \mbox{k}\Omega}^{-1}$. 
This was first established using semiconductor structures \cite{datta.meso95,imry.meso97,ferry.trans97} 
at $4$ K, but recent experiments on gold contacts have demonstrated it at room temperature \cite{hansen.prb97}.
How can a wire have a resistance that is independent of
its length? The answer is that this resistance is really associated with the interfaces between 
the narrow wire and the wide contacts. If there is scattering inside the wire it would give 
rise to an additional resistance in series with this fundamental interface resistance. 
The fact that a short wire has a resistance of $12.9$ k$\Omega$ is a non-obvious result 
that was not known before 1988. This is a problem for which we do not really need a
quantum transport formalism; a semi-classical treatment would suffice. The results we obtain here are not new
or surprising. What is new is that we treat the gold wire as an $Au_6$ molecule and obtain well known results
commonly obtained from a continuum treatment. 

In order to apply the NEGF formalism from the last section to this problem, we need the Hamiltonian matrix $[H]$,
the self energy matrices $\Sigma_{1,2}$ and the self-consistent field $U_{SCF}=F([\rho])$.
Let us look at these one by one.

\emph{Hamiltonian}: We will use a simple semi-empirical Hamiltonian which uses one s-orbital
centered at each gold atom as the basis functions with the elements of the Hamiltonian matrix given by:
\beqa
H_{ij}&= \epsilon_0 & \mbox{if} \; i=j \label{eq.hamiltonian} \\
&=-t&\mbox{if} \; i,j \; \mbox{are nearest neighbors} \nonumber
\eeqa
where $\epsilon_0=-10.92$ eV and $t=2.653$ eV. The orbitals are assumed to be orthogonal, 
so that the overlap matrix $S$ is the identity matrix.

\begin{figure}[!htb]
\begin{center}
\includegraphics[width=\myfigsized \columnwidth]{\myfigpatha/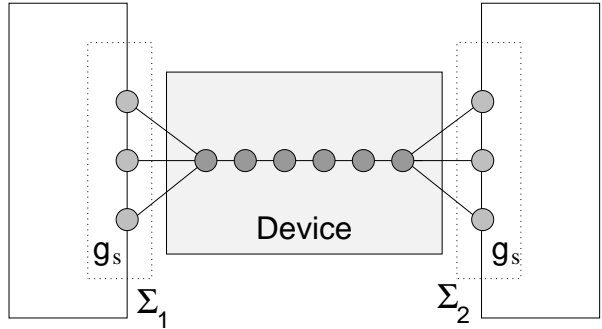}
\end{center}
\caption{Device, surface Green's function 
($g_s$) and Self-energies ($\Sigma$).}
\label{fig.overview}
\end{figure}

\emph{Self Energy}:
Once we have a Hamiltonian for the entire molecule-contact system,
the next step is to ``partition'' the device from the contacts and obtain 
the self-energy matrices $\Sigma_{1,2}$ describing the effects
of the contacts on the device.
The contact will be assumed to be essentially unperturbed relative to the 
surface of a bulk metal so that the full Green's function ($G_T$) can be
written as (the energy $E$ is assumed to have an infinitesimal imaginary part $i 0^+$):
\beq
G_T=\mymat{2}{E S-H& E S_{dc}-H_{dc}\\E S_{cd}-H_{cd} & E S_{c}-H_{c}}^{-1}=
\mymat{2}{G& G_{dc}\\ G_{cd}&  G_{C}} \label{eq.datta209}
\eeq
where ``$c$'' denotes one of the contacts (the effect of the other contact can be obtained separately).
We can use straight forward matrix algebra to show that:
\beqa
G&=&\mybpar{E S-H-\Sigma}^{-1}\\
\Sigma&=&\mybpar{E S_{dc}-H_{dc}}\mybpar{E S_c-H_c}^{-1}\mybpar{E S_{cd}-H_{cd}}
\eeqa
The matrices $S_{dc}$, $S_{c}$, $H_{dc}$, $H_{c}$ are all infinitely large since the contact is infinite. 
But the element of $S_{dc}$, $H_{dc}$ are non-zero only for a small number of contact atoms whose wavefunctions
significantly overlap the device. Thus, we can write:
\beq
\Sigma=\tau g_s \tau^\dagger \label{eq.dattaigen}
\eeq 
where $\tau$ is the non-zero part of $E S_{dc}-H_{dc}$ having dimensions ($d \times s$) where 
'$d$' is the size of the device matrix , and '$s$' is the number of surface atoms of the contact
having a non-zero overlap with the device. $g_s$ is a matrix of size $s \times s$  which is a
subset of the full infinite-sized contact Green's function $\mybpar{E S_c-H_c}^{-1}$.
This surface Green's function 
can be computed exactly by making use of the periodicity of the semi-infinite contact,
using techniques that are standard in surface physics \cite{yellow.book}.
For a one-dimensional lead, with a Hamiltonian given by Eq.~\ref{eq.hamiltonian},
the result is easily derived \cite{datta.meso95}:
\beq
g_s(E)=-\frac{e^{i k a}}{t} \label{eq.onedsgf}
\eeq
where ``$k a$'' is related to the energy through the dispersion relation:
\beq
E=\epsilon_0-2 t \; \mbox{cos}(k a)
\eeq
The results presented below were obtained using the more complicated surface Green's 
function for an FCC (111) gold surface as described in Ref.~\cite{zahid.mark02}.
However, using the surface Green's function in Eq.~\ref{eq.onedsgf} gives almost
identical results.

\begin{figure}[!htb]
\begin{center}
\includegraphics[width=\myfigsized \columnwidth]{\myfigpatha/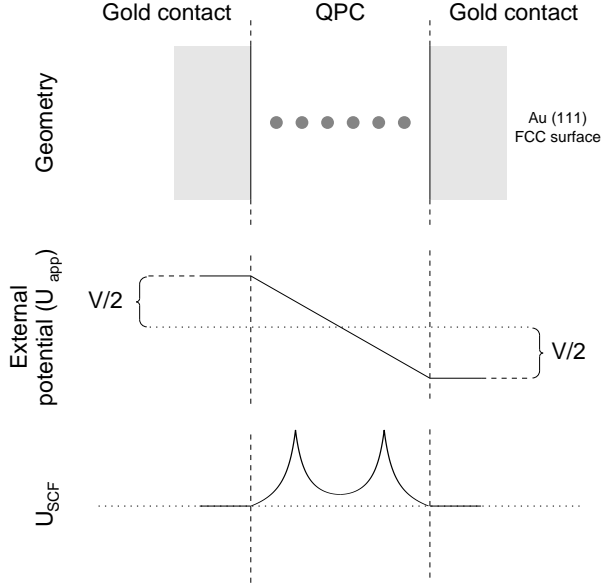}
\end{center}
\caption{The electrostatic potentials divided into the applied ($U_{app}$)
and self consistent field ($U_{SCF}$) potentials. 
The boundary conditions can clearly be seen in the figure, $U_{SCF}$ is zero at the boundary and
$U_{app}=\pm V_{app}/2$.}
\label{fig.estatic}
\end{figure}

\emph{Electrostatic potential}:
Finally we need to identify the electrostatic potential across the device (Fig.~\ref{fig.estatic})
by solving the Poisson equation:
\beq
- \nabla^2 U_{tot}= \frac{e^2 n}{\epsilon_0} \label{eq.poisson}
\eeq
with the boundary conditions given by the potential difference $V_{app}$ between the metallic contacts 
(here $\epsilon_0$ is the dielectric constant).
To simplify the calculations we divide the solution into a applied and self consistent potential 
($U_{tot}=U_{app}+U_{SCF}$) where $U_{app}$ solves the Laplace equation with the
known potential difference between the metallic contacts: 
\beqa
\nabla^2 U_{app}=0 & & U_{app}=-e V_n \; \mbox{on electrode 'n'} 
\eeqa
Thus, $U_{tot}$ solves Eq.~\ref{eq.poisson}
if $U_{SCF}$ solves Eq.~\ref{eq.poisson} with zero potential at the boundary.
\beqa
\nabla^2 U_{SCF}=-\frac{e^2 n}{\epsilon_0} & & U_{SCF}=0 \; \mbox{on all electrodes} 
\eeqa
In the treatment of the electrostatic we assume the two contacts to be semi-infinite classical metals
separated by a distance ($W$). This gives simple solutions to both $U_{app}$ and $U_{SCF}$. 
The applied potential is given by (capacitor):
\beq
U_{app}=\frac{V}{W} x
\eeq
where x is the position relative to the midpoint between the contacts.
The self consistent potential is easily calculated with the method of images where the potential is given by
a sum over the point charges and all their images.
However, to avoid the infinities associated with point charges, 
we adopt the Pariser-Parr-Pople (PPP) method \cite{murrell.semi72,paulsson.prb01} in the Hartree approximation.
The PPP functional describing the electron-electron interactions is:
\begin{equation}
H^{e-e}_{ij} = \delta_{ij} \sum_{k} 
\left(\rho_{kk}-\rho_{kk}^{eq}\right) \gamma_{ik}
\label{eq.e-e}
\end{equation}
where $\rho$ is the charge density matrix, $\rho^{eq}$ the equilibrium charge density (in this case 
$\rho^{eq}_{ii}=1$ since we are modeling the s-electrons of gold)
and the one center two-electron integral $\gamma_{ij}$. The diagonal elements 
$\gamma_{ii}$ are obtained
from experimental data and the off-diagonal elements ($\gamma_{ij}$) 
are parameterized to describe a potential
that decrease as the inverse of the distance ($1/R_{ij}$):  
\beq
\gamma_{ij}=\frac{e^2}{4 \pi \epsilon_0 R_{ij}+\frac{2 e^2}{\gamma_{ii}+\gamma_{
jj}}} \label{eq.ppp}
\eeq

\begin{figure}[!htb]
\begin{center}
\includegraphics[width=\myfigsizec \columnwidth]{\myfigpatha/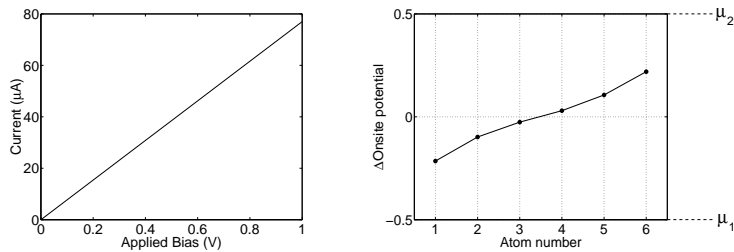}
\end{center}
\caption{I-V (left) and potential drop for an applied voltage of $1$ V (right) 
for a six atom QPC connected to two contacts. 
The potential plotted is the difference in onsite potential from the equilibrium case.} 
\label{fig.qpccurr}
\end{figure}

\emph{Calculations on the QPC}:
The results for the I-V and potential for a QPC are shown in Fig.~\ref{fig.qpccurr}.
The geometry used was a linear chain of six gold atoms connected to the FCC (111) surface of the contacts
in the center of a surface triangle. The Fermi energy of the isolated 
contacts was calculated to be $E_f=-8.67$ eV, by requiring that there is one electron per unit cell. 

As evident from the figure, the I-V characteristics is linear and the slope gives a conductance of 
$77.3 \; \mu \mbox{A}/\mbox{V}$ 
close to the quantized value of $\frac{e^2}{\pi \hbar}\sim 77.5 \; \mu \mbox{A}/\mbox{V}$ as previously mentioned.
What makes the QPC distinct from typical molecules is the strong coupling to the contacts which 
broadens all levels into a continuous density of states and any evidence of a conductance gap 
(Fig.~\ref{fig.i-v}) is completely lost.
Examining the potential drop over the QPC shows a linear drop over the center of the QPC with 
slightly larger drop at the end atoms. This may seem surprising since transport is
assumed to be ``ballistic'' and one expects no voltage drop across the chain of gold atoms.
This can be shown to arise because the chain is very narrow (one atom in cross-section) compared
to the screening length \cite{damle.prb01}.

\begin{figure}[!htb]
\begin{center}
\includegraphics[width=\myfigsizec \columnwidth]{\myfigpatha/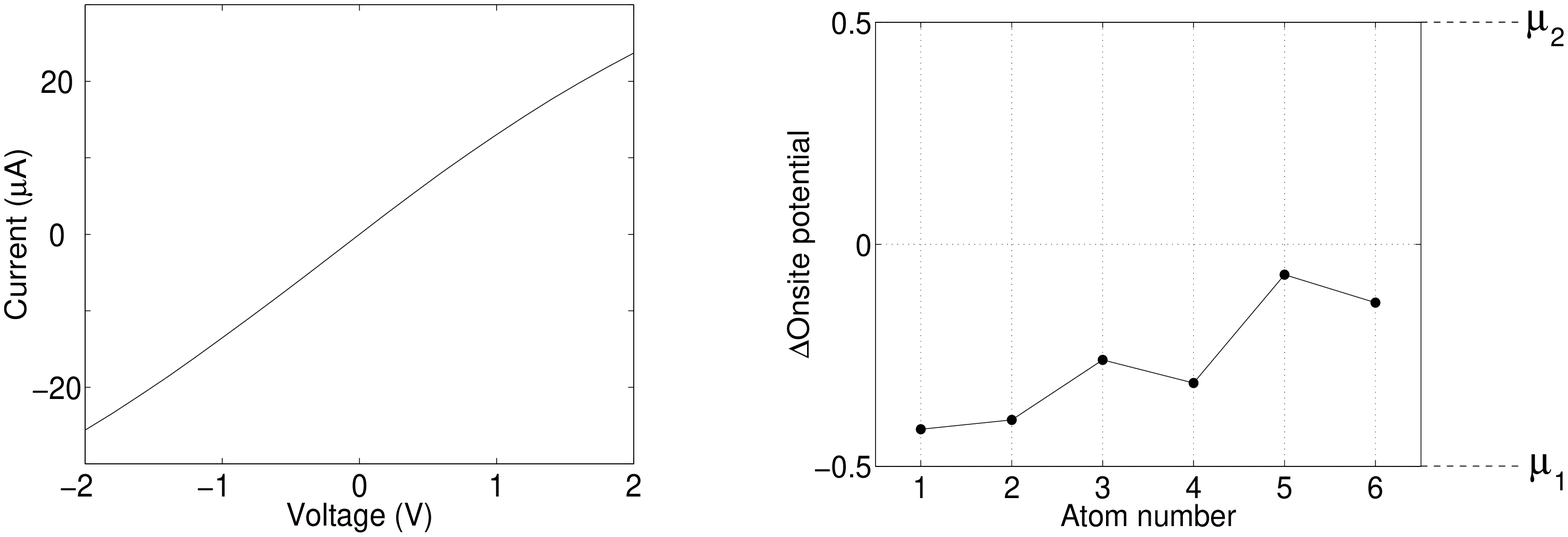}
\end{center}
\caption{I-V and potential drop for an applied voltage of $1$ V for the QPC asymmetrically 
connected the gold contacts.
The coupling to the right contact used is $-0.2 t$. } 
\label{fig.qpcasymcurr}
\end{figure}

We can easily imagine a experimental situation where the device (QPC or molecule) is attached
asymmetrically to the two contacts with one strong and one weak side. To model this situation
we artificially decreased the interaction between the right contact and the QPC by a factor of $0.2$.
The results of this calculation are shown in Fig.~\ref{fig.qpcasymcurr}. A weaker coupling gives a
smaller conductance, as compared with the previous figure. More interesting is that the potential
drop over the QPC is asymmetric. Also in line with our classical intuition, 
the largest part of the voltage drop occurs at the weakly coupled contact with smaller drops over the 
QPC and at the strongly coupled contact. The consequences of asymmetric voltage drop over molecules has 
been discussed by Ghosh {\it et al} \cite{avik.unpub01}.

\section{Concluding Remarks}

In this chapter we have presented an intuitive description of the current-voltage (I-V)
characteristics of molecules using simple toy models to illustrate the basic physics 
(sections \ref{sect.intro}-\ref{sect.charging}). 
These toy models were also used to motivate the rigorous Non-Equilibrium Green's 
Function (NEGF) theory (section \ref{sect.theory}). A simple example was then used in section \ref{sect.results}
to illustrate the application of the NEGF formalism. The same basic approach can be used 
in conjunction with a more elaborate H\"uckel Hamiltonian  
or even an ab initio Hamiltonian. But for these advanced treatments we refer the 
reader to Refs.~\cite{zahid.mark02,avik.chap02}.

Some of these models are publicly available through the Purdue Simulation Hub (www.\-nanohub.\-purdue.\-edu) 
and can be run without any need for installation. In addition to the models discussed here, 
there is a H\"uckel model which is an improved version of the earlier model made available in 1999.
Further improvements may be needed to take into account the role of inelastic scattering or polaronic effects, 
especially in longer molecules like DNA chains.

\section*{Acknowledgments}
Sections \ref{sect.quality} and \ref{sect.charging} are based on
material from a forthcoming book by one of the authors (S.D.) entitled ``Quantum Phenomena: 
From Atoms to Transistors''.
It is a pleasure to acknowledge helpful feedback from Mark Ratner, Phil Bagwell and Mark Lundstrom.
The authors are grateful to Prashant Damle and Avik Ghosh for helpful discussions regarding the 
ab initio models.
This work was supported by the NSF under 
Grant No. 0085516 - EEC. 

\appendix

\section{MATLAB Codes}
The Matlab codes for the toy models can also be obtained at ``www.nanohub.purdue.edu''.
\subsection{Discrete One Level Model}
\label{app.toy1}
\begin{verbatim}
%  Toy model, one level
%  Inputs (all in eV)
E0=-5.5;Ef=-5;gam1=0.2;gam2=0.2;U=1;
%  Constants (all MKS, except energy which is in eV)
hbar=1.06e-34;q=1.6e-19;IE=(2*q*q)/hbar;kT=.025;
%  Bias (calculate 101 voltage points in [-4 4] range)
nV=101;VV=linspace(-4,4,nV);dV=VV(2)-VV(1);
N0=2/(1+exp((E0-Ef)/kT));
for iV=1:nV %  Voltage loop
  UU=0;dU=1;
  V=VV(iV);mu1=Ef-(V/2);mu2=Ef+(V/2);
  while dU>1e-6 %  SCF
    E=E0+UU;
    f1=1/(1+exp((E-mu1)/kT));f2=1/(1+exp((E-mu2)/kT));
    NN=2*((gam1*f1)+(gam2*f2))/(gam1+gam2); %  Charge
    Uold=UU;UU=Uold+(.05*((U*(NN-N0))-Uold));
    dU=abs(UU-Uold);[V UU dU];
  end
  curr=IE*gam1*gam2*(f2-f1)/(gam1+gam2);
  II(iV)=curr;N(iV)=NN;[V NN];
end
G=diff(II)/dV;GG=[G(1) G]; %  Conductance
h=plot(VV,II*10^6,'k'); %  Plot I-V
\end{verbatim}

\subsection{Discrete Two Level Model}
\label{app.toy2}
\begin{verbatim}
%  Toy model, two levels
%  Inputs (all in eV)
Ef=-5;E0=[-5.5 -1.5];gam1=[.2 .2];gam2=[.2 .2];U=1*[1 1;1 1];
% Constants (all MKS, except energy which is in eV)
hbar=1.06e-34;q=1.6e-19;IE=(2*q*q)/hbar;kT=.025;
n0=2./(1+exp((E0-Ef)./kT));
nV=101;VV=linspace(-6,6,nV);dV=VV(2)-VV(1);Usc=0;
for iV=1:nV
    dU=1;
    V=VV(iV);mu1=Ef+(V/2);mu2=Ef-(V/2);
    while dU>1e-6
        E=E0+Usc;
        f1=1./(1+exp((E-mu1)./kT));f2=1./(1+exp((E-mu2)./kT));
        n=2*(((gam1.*f1)+(gam2.*f2))./(gam1+gam2));
        curr=IE*gam1.*gam2.*(f1-f2)./(gam1+gam2);
        Uold=Usc;Usc=Uold+(.1*(((n-n0)*U')-Uold));
        dU=abs(Usc-Uold);[V Usc dU];
    end
    II(iV)=sum(curr);N(iV,:)=n;
end
G=diff(II)/dV;GG=[G(1) G];
h=plot(VV,II); %  Plot I-V
\end{verbatim}

\subsection{Broadened One Level Model}
\label{app.toy3}
\begin{verbatim}
%  Toy model, restricted solution with broadening
%  Inputs (all in eV)
E0=-5.5;Ef=-5;gam1=0.2;gam2=0.2;U=1.0;
%  Constants (all MKS, except energy which is in eV)
hbar=1.06e-34;q=1.6e-19;IE=(2*q*q)/hbar;kT=.025;
%  Bias (calculate 101 voltage points in [-4 4] range)
nV=101;VV=linspace(-4,4,nV);dV=VV(2)-VV(1);
N0=2/(1+exp((E0-Ef)/kT));
for iV=1:nV %  Voltage loop
  UU=0;dU=1;
  V=VV(iV);mu1=Ef-(V/2);mu2=Ef+(V/2);
  nE=400; %  Nummerical integration over 200 points 
  id=diag(eye(nE))';
  EE=linspace(-10,0,nE);dE=EE(2)-EE(1);
  f1=1./(1+exp((EE-id*mu1)/kT));
  f2=1./(1+exp((EE-id*mu2)/kT));
  while dU>1e-4 %  SCF
    E=E0+UU;
    g=1./(EE-id*(E+i/2*(gam1+gam2)));
    NN=2*sum(g.*conj(g).*(gam1*f1+gam2*f2))/(2*pi)*dE;
    Uold=UU;UU=Uold+(.2*((U*(NN-N0))-Uold));
    dU=abs(UU-Uold);[V UU dU];
  end
  curr=IE*gam1*gam2*sum((f2-f1).*g.*conj(g))/(2*pi)*dE;
  II(iV)=real(curr);N(iV)=NN;[V NN curr E mu1 mu2];
end
G=diff(II)/dV;GG=[G(1) G]; %  Conductance
h=plot(VV,II,'.'); %  Plot I-V
\end{verbatim}

\subsection{Unrestricted Discrete One-Level Model}
\label{app.toy4}
\begin{verbatim}
% Toy model unrestricted solution
%  Inputs (all in eV)
E0=-5.5;Ef=-5;gam1=0.2;gam2=0.2;U=1;
%Constants (all MKS, except energy which is in eV)
hbar=1.06e-34;q=1.6e-19;IE=(q*q)/hbar;kT=.025;
%  Bias (calculate 101 voltage points in [-4 4] range)
nV=101;VV=linspace(-4,4,nV);dV=VV(2)-VV(1);
N0=1/(1+exp((E0-Ef)/kT));
for iV=1:nV %  Voltage loop
  U1=0;U2=1e-5;dU1=1;dU2=1;% Set U2=U1 for restricted solution
  V=VV(iV);mu1=Ef-(V/2);mu2=Ef+(V/2);
  while (dU1+dU2)>1e-6 %  SCF
    E1=E0+U1;E2=E0+U2;
    f11=1/(1+exp((E1-mu1)/kT));f21=1/(1+exp((E1-mu2)/kT));
    f12=1/(1+exp((E2-mu1)/kT));f22=1/(1+exp((E2-mu2)/kT));
    NN1=((gam1*f12)+(gam2*f22))/(gam1+gam2);
    NN2=((gam1*f11)+(gam2*f21))/(gam1+gam2);
    Uold1=U1;Uold2=U2;
    U1=Uold1+(.05*((2*U*(NN1-N0))-Uold1));
    U2=Uold2+(.05*((2*U*(NN2-N0))-Uold2));
    dU1=abs(U1-Uold1);dU2=abs(U2-Uold2);
  end
  curr1=IE*gam1*gam2*(f21-f11)/(gam1+gam2);
  curr2=IE*gam1*gam2*(f22-f12)/(gam1+gam2);
  I1(iV)=curr1;I2(iV)=curr2;
  N1(iV)=NN1;N2(iV)=NN2;[V NN1 NN2];
end
G=diff(I1+I2)/dV;GG=[G(1) G]; %  Conductance
h=plot(VV,I1+I2,'-'); %  Plot I-V
\end{verbatim}

\subsection{Unrestricted Broadened One-Level Model}
\label{app.toy5}
\begin{verbatim}
%  Toy model, unrestricted solution with broadening
%  Inputs (all in eV)
E0=-5.5;Ef=-5;gam1=0.2;gam2=0.2;U=1;
%  Constants (all MKS, except energy which is in eV)
hbar=1.06e-34;q=1.6e-19;IE=(q*q)/hbar;kT=.025;
%  Bias (calculate 101 voltage points in [-4 4] range)
nV=101;VV=linspace(-4,4,nV);dV=VV(2)-VV(1);
N0=1/(1+exp((E0-Ef)/kT));
nE=200; %  Nummerical integration over 200 points 
id=diag(eye(nE))';
EE=linspace(-9,-1,nE);dE=EE(2)-EE(1);
for iV=1:nV %  Voltage loop
  U1=0;U2=1;dU1=1;dU2=1;
  V=VV(iV);mu1=Ef-(V/2);mu2=Ef+(V/2);
  f1=1./(1+exp((EE-id*mu1)/kT));
  f2=1./(1+exp((EE-id*mu2)/kT));
  while (dU1+dU2)>1e-3 %  SCF
    E1=E0+U1;E2=E0+U2;
    g1=1./(EE-id*(E1+i/2*(gam1+gam2)));
    g2=1./(EE-id*(E2+i/2*(gam1+gam2)));
    NN1=sum(g1.*conj(g1).*(gam1*f1+gam2*f2))/(2*pi)*dE;
    NN2=sum(g2.*conj(g2).*(gam1*f1+gam2*f2))/(2*pi)*dE;
    Uold1=U1;Uold2=U2;
    U1=Uold1+(.2*((2*U*(NN2-N0))-Uold1));
    U2=Uold2+(.2*((2*U*(NN1-N0))-Uold2));
    dU1=abs(U1-2*U*(NN2-N0));dU2=abs(U2-2*U*(NN1-N0));
  end
  curr=IE*gam1*gam2*sum((f2-f1).*(g1.*conj(g1)+ ...
        g2.*conj(g2)))/(2*pi)*dE;
  II(iV)=real(curr);N(iV)=NN1+NN2;
  [V NN1 NN2 curr*1e6 E1 E2 mu1 mu2];
end
G=diff(II)/dV;GG=[G(1) G]; %  Conductance
h=plot(VV,II,'--'); %  Plot I-V
\end{verbatim}

\clearpage

\end{document}